\documentclass{article}
\input epsf.sty
\input colordvi.sty

\newcommand{\half}{\frac{1}{2}}
\newcommand{\bright}{\begin{flushright}}
\newcommand{\eright}{\end{flushright}}
\newcommand{\bminip}{\begin{minipage}}
\newcommand{\eminip}{\end{minipage}}
\newcommand{\bcent}{\begin{center}}
\newcommand{\ecent}{\end{center}}
\newcommand{\beq}{\begin{equation}}
\newcommand{\eeq}{\end{equation}}
\newcommand{\beqa}{\begin{eqnarray}}
\newcommand{\eeqa}{\end{eqnarray}}
\newcommand{\barr}{\begin{array}}
\newcommand{\earr}{\end{array}}
\newcommand{\bfig}{\begin{figure}}
\newcommand{\efig}{\end{figure}}
\newcommand{\nnb}{\nonumber}
\newcommand{\reflef}{(\ref}
\newcommand{\MP}{M_{\rm P}}
\newcommand{\lmd}{\lambda}
\newcommand{\Lmd}{\Lambda}

\newcommand{\lsim}{\mbox{\raisebox{-.3em}{$\;\stackrel{<}{\sim}\;$}}}

\begin{document}
\baselineskip=0.6cm

\renewcommand{\thefootnote}{\fnsymbol{footnote}}
\bcent
{\LARGE\bf Accelerating universe, WEP violation and antihydrogen
atoms\footnote{Prepared for the Proceedings of the Workshop on Cold Antimatter Plasmas and Application to Fundamental Physics, February 20-22, 2008, Naha, Okinawa, Japan.  Due to a technical problem, diagrams in Fig. 4 in the original Proceedings are replaced here by simplified versions. }
}\\[.8em]
{\large Yasunori Fujii}\\[.6em]
Advanced Research Center for Science and Engineering, Waseda University, \\[-.4em]Tokyo,  169-8555, Japan
\ecent
\mbox{}\\[-3.5em]

\renewcommand{\thefootnote}{\arabic{footnote}}
\setcounter{footnote}{0}
\bcent
\bminip{13cm}
\begin{flushleft}
{\large\bf Abstract}\\[1.0em]
\end{flushleft}

Apart from the suspected violation of the CPT invariance, we might expect if the measurements of antihydrogen atoms provide testing  Weak Equivalence Principle (WEP) in the gravitational phenomena.  We start with how its violation can be related to the expected idea of unification of particle physics and gravitation, an attempt beyond the standard theories, including general relativity of Einstein.  A particular emphasis will be placed on the issue of an accelerating universe, a rather recent development since nearly 10 years ago, suggesting a strong motivation toward attempts beyond the conventional concepts of the traditional cosmology.  We face today's version of the cosmological constant problem.  A candidate of the new ingredient appears to be provided by a scalar field, sometimes under the names like quintessence or dark-energy.  In this article, we discuss the subject from a point of view of a more theoretical approach based on the scalar-tensor theory of gravitation.  By exploiting the concepts of conformal transformations and conformal frames unique to this type of approach, we show that a successful understanding of the observed cosmological acceleration entails an unexpected outcome of breakdown of WEP, which may show up as a distinct behavior between  hydrogen and antihydrogen in the gravitational influence, from a further wider perspective including a vector field as well.  We intended to introduce most of the new concepts as plainly and briefly as possible, according to the nature of the talk.   On the other hand, however, we find it also useful to go into some more details, depending on the interests of individuals among the audience.  For a compromise, we decided to add a number of footnotes which were not delivered during the presentation, but might be helpful in the proceedings published later.

\eminip
\ecent

\section{Introduction}


From the point of view of general relativity (GR), we have no distinction  between particles and antiparticles in any of the gravitational phenomena.  But on the other hand, there has been a strong quest for unifying  particle physics and gravity.  In this approach, we might be allowed to make the distinction mentioned above.  In fact in almost any of the theoretical models of unification we find a number of fields which are yet to be discovered, but supposed to couple to the ordinary type of matter nearly as weakly as the ordinary gravity does.  For this reason we call them Non Geometric Gravitational Fields (NGGF), partially under the name GF.  At the same time, however, we find that these fields  belong to a category of the matter in Einstein's sense, instead of another category that accommodates the metric field.  In order to remind the audience of this non-geometric nature, we decided to put another attribute NG.  Due to this feature, these fields or forces will couple to the matter depending on some specific properties of individual object.  Consequently, matter objects will fall freely with acceleration different from object to object, hence violating the law of universal free-fall (UFF).\footnote{Basically Galilei's observation in the Leaning Tower in Pisa.}  This is also called a violation of Weak Equivalence Principle (WEP).  What do we mean by the term ``weak"?  We offer a very brief reply.

WEP represents a phenomenological aspect of the more general law expressed in the mathematical term; a tangential spacetime attached to any point on the Riemannian manifold is Minkowskian.  The message is that, even with violation of WEP, we can still maintain Einstein's central idea of dynamics of spacetime geometry unspoiled.\footnote{Comma-goes-to-semicolon rule remains true \cite{mtw}, for example.}  The only thing is that what we call ``gravitational phenomena" might be extended to be contaminated partially by NGGF,\footnote{In the same way as the ordinary ``matter" fields, like the electromagnetic or the strong-interaction field, do not respect the universal coupling to other fields, like the electrons and nucleons, NGGFs do not observe the simple results as implied by UFF. } hopefully as one of the barely visible traces of the unification mechanism, supposed to be realized fully only at extremely high energies.   There can be many different kinds of NGGF.  In practice we may focus upon scalar or vector fields supposed to have a very small masses, sometimes also called non-Newtonian gravity \cite{yfnatap}, or the fifth force \cite{fifth}.

\section{Why WEP violation in cosmology?}
\subsection{Accelerating universe}

Let us now tell how the cosmology comes to suggesting WEP violation.  As usual, we assume a homogeneous and isotropic universe, with $\rho (t)$, the matter density depending only on the cosmic time $t$.  Typical inter-galactic separations grow in proportion to the scale factor $a(t)$, which obeys the Einstein equation,
\beq
3H^2=\rho,\quad \mbox{with}\quad H=\frac{\dot{a}}{a}.
\label{2_1}
\eeq
In front of the density $\rho$ we usually have the coefficient like $8\pi G/c^2$.  But we use the (reduced) Planckian unit system with 
\beq
c=\hbar=\MP(=(8\pi G/(c\hbar))^{-1/2})=1. 
\label{2_2}
\eeq
The units of length, time and energy are re-expressed as follows in the traditional units,
\beq
8.10\times 10^{-33}{\rm cm}, \; 2.70\times 10^{-43}{\rm s};\;  2.44\times 10^{18}{\rm GeV},\quad (t_0=13.7 {\rm Gy}\approx 10^{60.2}).
\label{2_3}
\eeq

Also the Planck mass $\MP$ defined  in \reflef{2_2}) in terms of Newton's constant $G$, or the gravitational constant, is supposed to provide the basic energy scale of the unification processes.   Particularly interesting is $t_0$, today's age, now believed to be 13.7 Gy.   In units of the Planck time, as shown in  the last portion of \reflef{2_3}), this is $\sim 10^{60}$, a number worthwhile to be remembered, as we are going to show shortly.

Before 1998, we had taken it for granted that the universe was decelerated, because under virtually any of the reasonable assumptions, we always ended up with $a(t)$ which behaves in the left diagram in Fig. 1, with $a''(t)<0$, thus decelerating.  In 1998, however, two groups of astronomers reported  their detailed observation of the very distant objects, with a startling result;  an accelerating universe with $a''(t)>0$, as shown by the right diagram \cite{accel}.

\mbox{}\\[-19.0em]
\bminip[t]{12cm}
\hspace*{5.5em}
\epsfxsize=10.0cm
\epsffile{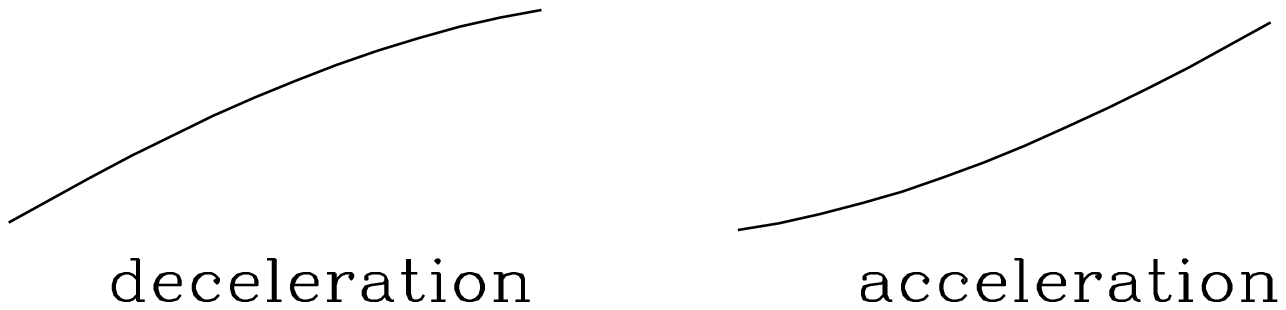}
\mbox{}\\[-2.5em]
\hspace*{10.2em}
Figure. 1: Two typical behaviors of $a(t)$.
\eminip
\mbox{}\\[0.0em]

They also showed that they fitted their observation by adding a positive constant $\Lmd >0$ on the right-hand side of the Einstein equation;
\beq
3H^2=\rho+\Lmd.
\label{2_4}
\eeq
We add an explanation based on Fig. 2 which was prepared for the sake of illustration, in which we conveniently introduced the ``critical density" $\rho_{\rm cr}$ defined by
\beq
\rho_{\rm cr}\equiv 3H_0^2,
\label{2_5}
\eeq
implying, with the subscript 0 for the present time, the entire right-hand side  of \reflef{2_4}).\\[-4.1em]

\hspace*{3.8em}
\bminip{12cm}
\hspace*{4.3em}
\epsfxsize=8.0cm
\epsffile{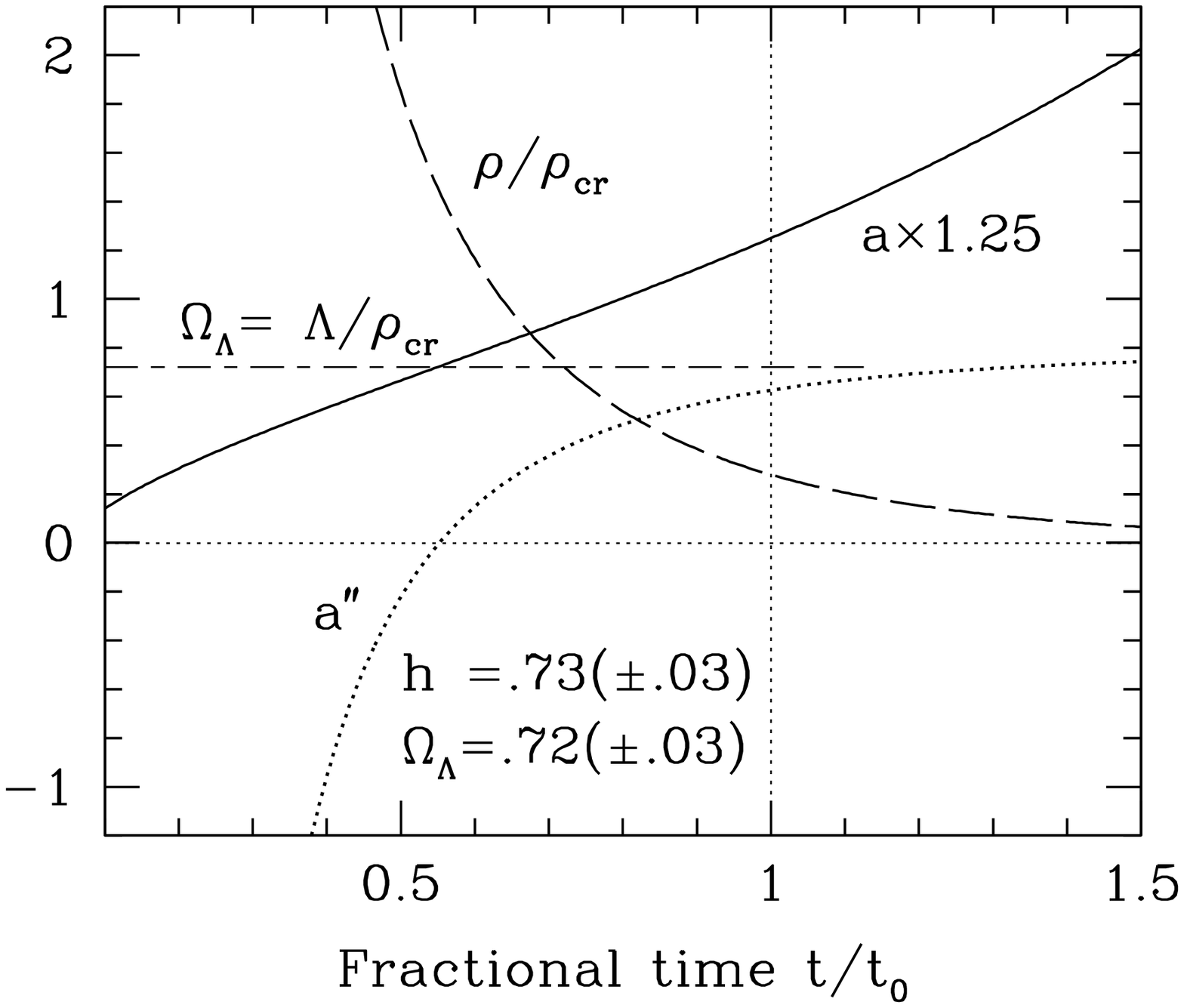}
\mbox{}\\[-.1em]
\hspace*{4em}Figure 2: Illustrative plots of $\rho(t), a(t), a''(t), \Lmd$ against $t/t_0$.\eminip
\mbox{}\\[1.2em]

As we find, $\rho$ or $\rho/\rho_{\rm cr}$ falls off generically like $t^{-2}$, while $\Lmd$ or $\Omega_\Lmd\equiv \Lmd/\rho_{\rm cr}$ stays constant.  They cross each other.  Before this crossing, $a$ is decelerating, while after the crossing, $a$ is accelerating, though slightly, roughly speaking.  The observation by these groups determined $\Omega_\Lmd \sim 0.7$.  This implies that so far unknown component $\Lmd$ makes up as much as 70\% of the entire cosmic energy.  In the Planckian unit system, $\rho_{\rm cr}$ is given basically like $t_0^{-2}$.  As was shown in \reflef{2_3}), we thus find 
\beq
\Lmd_{\rm obs}\approx\rho_{\rm cr} \approx (10^{60})^{-2}\approx 10^{-120},
\label{2_6}
\eeq
where we used the result $\Omega_\Lmd \approx 1$ for the ``observed" value $\Lmd_{\rm obs}$.

On the other hand, nearly any of the theoretical models of unification showed it unavoidable to have $\Lmd$ of the order of $\sim \MP^4$ or $\sim 1$ in the reduced Planckian unit system.  This implies $\Lmd_{\rm th}\sim 1$, which we compare with the observed value.  This is certainly the worst record of disagreement between theory and observation.  This is today's version of the cosmological constant problem, which consists of the two questions.

First, what a powerful theory is it to make it possible to fine-tune the parameters to the incredible accuracy of 120 orders of magnitude? This is a fine-tuning problem.  Suppose we may nevertheless assume a truly constant $\Lmd$ of this small.  As we showed in Fig. 2, it crosses $\rho$, hence causing an acceleration only once throughout the whole history of the universe, which we are now watching in some sense.  What a coincidence?  This is the coincidence problem.

From these considerations we come to wondering if $\Lmd$ is not a true constant, but is of some dynamical nature.  In this connection we notice that the $\Lmd$ term on the right-hand side of \reflef{2_4})  can also be described by an ideal fluid with the equation of state $p=-\Lmd <0$.  It is this negative pressure that provides an extra acceleration to the universe.  Also this negative pressure is precisely what a scalar field distributed uniformly everywhere in the universe, as described by $\phi (t)$, can provide with.  This is sometimes called a dark energy different from the dark matter.

Some people wish to see the scalar field as a major player in the universe under the name of quintessence \cite{quint}, but following quite a cautious and phenomenological approach.  We are going to say, however, that a more theoretical approach could be not only promising but also rewarding.

\subsection{Scalar-tensor theory}

As one of such theoretical approaches, we are going to talk about the scalar-tensor theory, invented in 1955 by Jordan \cite{jord} with the basic Lagrangian 
\beq
{\cal L}_{\rm ST\Lmd}= \sqrt{-g} \left( \half \xi \phi^2 R -\epsilon \half g^{\mu\nu}\partial_\mu\phi \partial_\nu\phi +L_{\rm matter} -\Lmd \right),
\label{2_7}
\eeq
where $\phi$ is the scalar field, with its kinetic energy term, the second term.  We use  the notation slightly different from Jordan's $\omega$, but nothing is serious about this.\footnote{The exact relations to Jordan's original notations $\phi_{\rm J}$ and $\omega$ are given by $16\pi \phi_{\rm J}=(\xi/2)\phi^2, \;\epsilon ={\rm Sign}(\omega), \;\xi^{-1}=4|\omega |$.}  We use again the Planckian unit system.  Unlike in Jordan's original theory, we have added $\Lmd$ in order to show the relevance to the cosmological constant problem.

The first term on the right-hand side of \reflef{2_7}) is called a nonminimal coupling (NM) term, which is truly unique to this theory, accommodating $\phi$ in a nontrivial manner.  We compare this with the traditional Einstein-Hilbert (EH) term in GR;
\beq
{\cal L}_{\rm EH}=\sqrt{-g}\frac{1}{16\pi G}R=\sqrt{-g}\half R,
\label{2_8}
\eeq
showing that this theory has no truly constant $G$.  Instead it has an effective $G_{\rm eff}$ defined by 
\beq
\frac{1}{8\pi G_{\rm eff}}= \xi\phi^2(x),
\label{2_9}
\eeq
in terms of the spacetime function $\phi(x)$.  One of Jordan's motivations was to propose a consistent theory which allows a time-dependent gravitational ``constant,"  as had been suggested earlier by Dirac \cite{dirac}.

Sometime later Brans and Dicke added an important feature \cite{bd}.  They required that $\phi$ is decoupled from $L_{\rm matter}$, or $\phi$ does not enter $L_{\rm matter}$.  By doing so they save the idea of WEP. The derivation in this ``Brans-Dicke model" is a little complicated.  For more technical details, we would like to ask the reader to refer to our book \cite{cup}, or the related paper \cite{yfinv}.

We also face another complication related to what is known as  a conformal transformation
\beq
g_{\mu\nu} \rightarrow g_{*\mu\nu} =\Omega^2(x) g_{\mu\nu},
\label{2_10}
\eeq
with an arbitrary spacetime function $\Omega (x)$.  This is a local scale transformation of the metric.  The Lagrangian in \reflef{2_7}), for example, is a specific functional of $g_{\mu\nu}$.  We substitute the inverse of \reflef{2_10}) into \reflef{2_7}), hence re-expressing the same Lagrangian as a different functional of the new metric $g_{*\mu\nu}$.  In  a sense the same Lagrangian looks different.  This might be termed as different Lagrangescapes \cite{yfinv}.

Pauli pointed out that the importance of a special choice
\beq
\Omega =\xi^{1/2}\phi,
\label{2_11}
\eeq
for which the re-expressed Lagrangian is put into
\beq
{\cal L}_{\rm ST\Lmd}\hspace{-.1em}=\hspace{-.2em}\sqrt{-g_{*}}\!\left(\half R_{*}\! - \!\half 	g^{\mu\nu}_{*}\partial_{\mu}\sigma\partial_{\nu}\sigma   +\!L_{\rm *{\rm matter}}\! - V(\sigma) \right), 
\label{2_12}
\eeq
with the EH term, instead of the NM term.  For this reason we say that by this transformation we have moved to the Einstein conformal frame (ECF) from the Jordan conformal frame (JCF).  Because $\Omega$ is arbitrary, we have infinitely many different CFs.  Out of them the above two, JCF and ECF, are of distinctive importance among others.

We notice that the ``canonical" scalar field in \reflef{2_12}) is no longer the original $\phi$ but is $\sigma$ newly defined by
\beq
\phi=\xi^{-1/2}e^{\zeta\sigma},
\label{2_13}
\eeq
where $\zeta$ is given by $\zeta^{-2}=6+\epsilon \xi^{-1}=6+4\omega$.  Also important is to note that the $\Lmd$ term in \reflef{2_7}) has been converted to the potential of the scalar field $\sigma$,
\beq
V=\Lmd e^{-4\zeta\sigma}.
\label{2_14}
\eeq

Now $G$ in \reflef{2_7}) is variable while the gravitational constant, which might be denoted by $G_*$, in \reflef{2_12}) is a constant.  In other words, $G$ is constant or variable depending on the choice of the CF.  The same is true not only for $G$, but also for many other so-called constants, like particle masses, and the coupling constants.  It is then crucially important to determine what or which CF we are now in, or what or which CF a ``physical" CF is.  In order to reply this question, we have to go through a long series of detailed discussions.  Skipping all the details, we will show what our conclusion is like.\footnote{Chapter 3, also in Chapters 2.6 and 4.4, in \cite{cup} have been devoted to discussing various aspects of the concepts of conformal transformations and conformal frames.  The content and its consequences are restated, sometimes even extended, also in our recent paper \cite{yfinv}.}  
\begin{itemize}
\item JCF is a theoretical CF with a large $\Lmd \sim 1$ expected from unification,
\item ECF is a physical CF with a small observed $\Lmd_{\rm eff}$ as small as $10^{-120}$.
\end{itemize}

We add a few comments.  For $\Lmd$ in JCF, we assume the ``large" value $\sim \MP^4 \sim 1$ in order to keep a contact with the unification idea.  Sometimes this CF is called a string CF.  On the other  hand, we have ``small" and observed $\Lmd$ in ECF.  Let us explain why.

Consider the energy density $\rho_\sigma$ of the scalar field $\sigma$ in ECF; 
\beq
\rho_\sigma =\half \dot{\sigma}^2 + V(\sigma), \quad V=\Lmd e^{-4\zeta\sigma},
\label{2_15}
\eeq
where \reflef{2_14}) has been used, while the dot implies a differentiation with respect to the cosmic time $t_*$ in ECF.  We may interpret this $\rho_\sigma$ as an effective cosmological constant, or the dark energy.

According to the solution of the Einstein equation in ECF, we find,\footnote{See (4.106) of \cite{cup}. The same result had been obtained \cite{yfpr}, however, based on a somewhat complicated model.}
\beq
\rho_\sigma \sim t_*^{-2}.
\label{2_16}
\eeq
At the present time this is
\beq
\sim t_{*0}^{-2}\sim (10^{-60})^2 = 10^{-120}.
\label{2_17}
\eeq
This is precisely the number in \reflef{2_6}) we found for $\Lmd_{\rm obs}$ to fit the data of acceleration.  This is nice.   This relation can be called a ``scenario of a decaying cosmological constant" which provides us with a simple and natural derivation of the mysterious number $10^{-120}$.  According to this derivation, today's $\Lmd$ is this small only because our universe is this old, not due to any of the fine-tuning of parameters, thus solving the fine-tuning problem as emphasized in Section 2.1. We would say that this is a major success of the scalar-tensor theory, not shared by the more phenomenology-oriented approaches, including the quintessence approach.\footnote{Strictly speaking, the simple falling-off time-dependence shown in \reflef{2_16}) is short of the role of a constant-like behavior which we should expect as a substitute of the cosmological constant causing an extra acceleration of the universe.  This issue demands a departure from a simple-minded scalar-tensor theory.  As discussed in detail in Chapter 5.4.2 of \cite{cup}, we introduced another scalar field $\chi$ to reproduce the occurrence of a temporarily constant-like behavior, whose size is basically given by \reflef{2_17}), inheriting the scenario of a decaying cosmological constant implemented before $\chi$ is introduced.  In this way we come also to replying the coincidence problem, at least by lessening its gravity.}

However, we still have a problem.  We started off with the matter Lagrangian according to the Brans-Dicke model in JCF.  Naturally the electron field, for example, as well as the electron mass term are present in $L_{\rm matter}$.  The absence of $\phi$ as required by WEP implies no time-dependence of the electron mass, $m_{\rm e}$.  Now a constant $m_{\rm e}$ in JCF is likely to be time-dependent in ECF, supposed to be a physical CF.  According to a more detailed analysis we do find a behavior;\footnote{See (4.121) of \cite{cup} and (3.13) of \cite{yfinv}.}
\beq
m_* = m t_*^{-1/2}\sim t_*^{-1/2}.
\label{2_18}
\eeq
This is too much time-dependent, as   we are going to argue in what follows.

Suppose we use atomic clocks in our measurements in physics, or in astronomy. Our time-unit  is provided basically by the electron mass.  Also suppose the electron mass changes with time, like \reflef{2_18}), for example.  Our time unit also does.  How can we detect that?  We can do it by comparing our clock with other kind of clock of a different principle.  Putting the other way around, we have no way to detect it, as long as we insist to use the atomic clock.  This might be stated as ``Own-unit insensitivity principle" \cite{yfinv}.  This might be translated into saying that $m_{\rm e}$ is a true constant in the physical CF which corresponds to the use of atomic clocks, whose time unit is provided by the same $m_{\rm e}$.  In other words, the physical CF should be such that it leaves $m_{\rm e}$ constant.  More generally speaking, choosing a physical CF depends on what clock we use.  From this point of view, the time-dependence like \reflef{2_18}) is unacceptable.

We may readily extend the same analysis to more realistic situation of measuring redshift of the light coming from far-away objects, hence estimating how long ago the light was emitted.  Eventually we come again to appreciating the constancy of the electron mass, in the same way as in using atomic clocks.\footnote{See  Section 3.3 of \cite{yfinv}.}

We emphasize again that choosing  ECF provides with many features that appear to qualify ECF to be a physical CF.  One of such features is the scenario of a decaying cosmological constant as mentioned above.

This argument above is a consequence of a long series of details, but is truly unique to the presence of $\Lmd$.  We would say that this is something hard to refute.  Then we have no other choice but to revise one of the initial assumptions that particle masses are constant in the JCF. We have to revise the Brans-Dicke model.  We in fact implemented the desired modification, resulting in the constant mass in ECF rather than in JCF.\footnote{See Chapter 4.4.3 of \cite{cup}.}  Obviously in the modified model back in JCF, $\phi$ enters the matter Lagrangian, against Brans-Dicke's premise, hence breaking WEP.  This is the way how we come to suspecting the validity of WEP,\footnote{The similar conclusion of WEP violation has been reached from string theory, but in a less compelling manner.} in general situations with the universe accelerating.\footnote{We admit that this conclusion derives from the scalar-tensor theory, part of which is too intricate to allow a naive interpretation. It is this theory, however, that, in spite of possible loose ends \cite{yfptp}, offers a theoretical implementation of the relation $\Lmd_{\rm obs}\sim 10^{-120}\sim (10^{60})^{-2}\sim t_0^{-2}$, hardly to be dismissed as a mere numerical coincidence.}

\section{Possible measurements on $\bar{\rm H}$}

We have specifically discussed the cosmological implications of the scalar field, which, we still believe, has an origin in the unification physics.  From this wider view, there is no reason why we confine ourselves only to the scalar field.  We may think of a vector field as well.  This choice is particularly interesting because it does allow a distinction between H and $\bar{\rm H}$.\footnote{In the language of the geodesic equation, NGGF contributes to the ``right-hand side." }  In what follows we will try two phenomenological attempts on the basis of the scalar and vector fields.

\subsection{Level-spacings of ${\rm H}$ and $\bar{\rm H}$}

We first try to see how much these fields affect the spectroscopy of H and $\bar{\rm H}$, like the energy levels between $1S$ and $2S$ with hyper-fine splittings included.  Let us begin with preparing a theoretical and artificial source of NGGF; a sphere with the radius $R$ filled with the substance of $\rho =10$gm/c.c., as an example.  We choose $R=10$m smaller than the force-range $\lmd$, supposed to be roughly around $100\mbox{m}$.\footnote{See (6.140) of \cite{cup}.}   We place the atoms near the surface with the $z$ axis chosen as shown in Fig. 3, with the origin at the center of $p$ or $\bar{p}$. \\[.8em]
\bminip{15cm}
\mbox{}\\[-.2em]
\hspace*{15.2em}
\epsfxsize=3.1cm
\epsffile{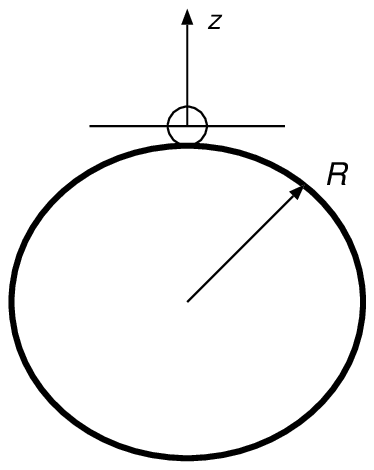}
\mbox{}\\[.5em]
\hspace*{5em}Figure 3: An artificial source of NGGF. $R=10\mbox{m}<\lmd \sim 100\mbox{m}, \hspace{.4em }\rho=10\mbox{gr}/\mbox{cc}$.

\eminip
\mbox{}\\[1.2em]

We are interested only in the order-of-magnitude estimate of the level energies.  Then we may ignore $p$ or $\bar{p}$ at the center.  In the non-relativistic approximation, we may also ignore the possible differences between $e$ and $\bar{e}$, and then between the scalar and vector fields.  In this context, the scalar field and the zeroth component of the vector field may be represented collectively by a single scalar field $\Phi$, which is produced by the artificial source as described above.  We also expand as $\Phi(z)\approx \Phi(0) + z\Phi'(0)$, with the coefficients evaluated as\\[-.9em]
\beqa
\Phi(0)&\sim&\zeta\int d^3r\frac{\rho e^{-r/\lmd}}{4\pi r} \sim \zeta\frac{M}{4\pi R} \sim \frac{1}{3}\zeta\rho R^2,\nnb\\
\Phi'(0) &\sim& -\Phi(0)/R. 
\label{3_1}
\eeqa

We try to import $\zeta$ from the diagrams in Fig. 4, which had been
obtained from various different sources.  In this connection, however,
we point out  that the value of $\zeta$, representing the coupling
strength of NGGF to the matter, is model-dependent.\footnote{A simple
relation $\zeta^2=\alpha_5/2$ can be derived only for the composition-independent experiments, i.e. probing possible deviation from the pure $1/r$ behavior, for the assumed form $V(r)=-(Gm_1m_2)(1+\alpha_5 e^{-r/\lmd})$ based on the Brans-Dicke model for the scalar field.  More uncertainties are unavoidable for any other kinds of estimates.}  They can be different for the scalar and the vector fields, for leptons and nucleons, and so on.  At present, we must be satisfied only by the crude and averaged estimates.  As such we tentatively choose $\zeta^2 \lsim 10^{-4}$.

\mbox{}\\[-5.0em]
\hspace*{2.5em}
\bminip{7.5cm}
\mbox{}\\[-.2em]
\hspace*{-1.2em}
\epsfxsize=6.5cm
\epsffile{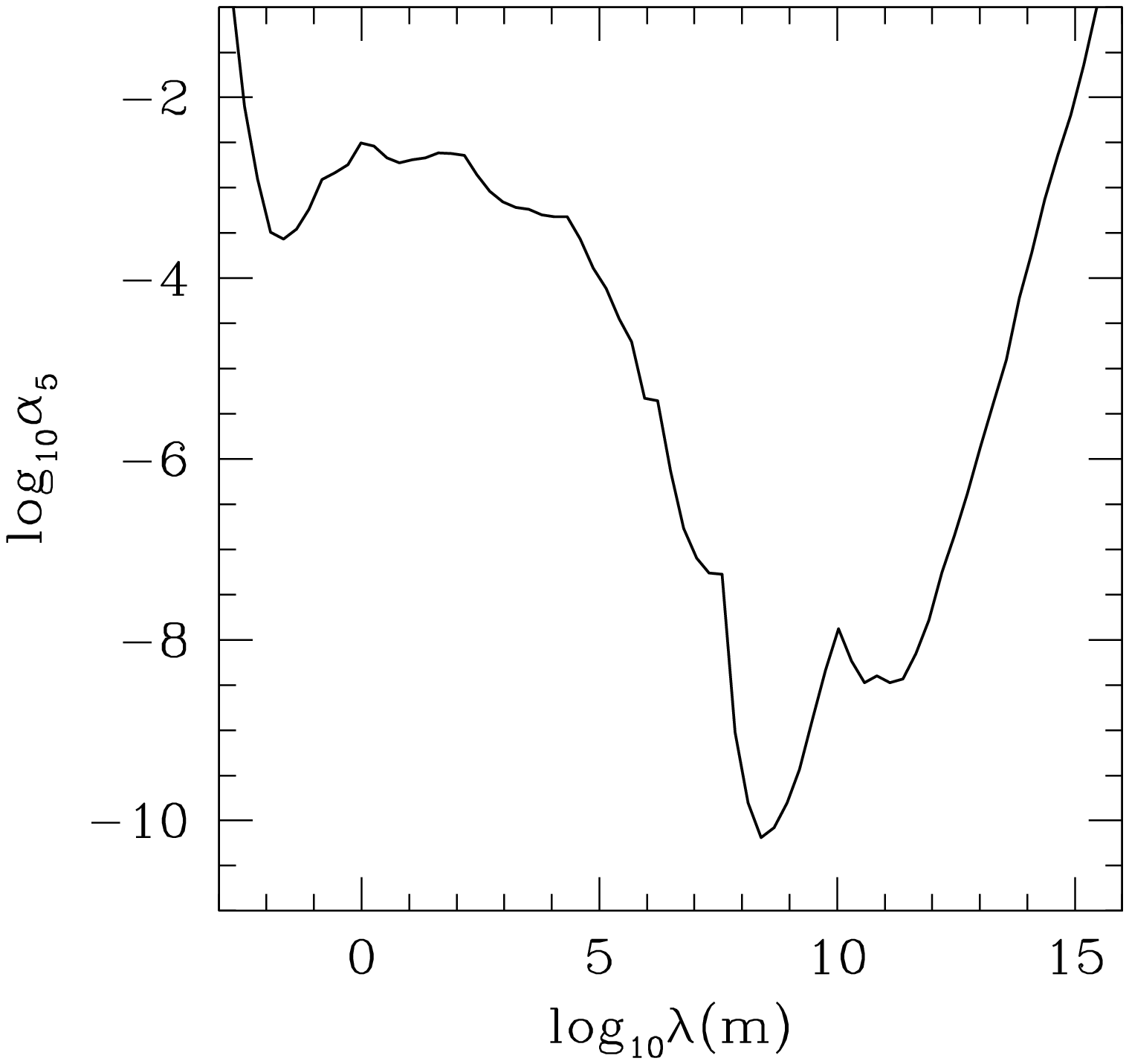}
\mbox{}\\[-.15em]
\hspace*{1.em}{\small Composition-independent experiments}
\eminip
\hspace{-1.5em}
\bminip{7.5cm}
\mbox{}\\[-.5em]
\hspace*{-.2em}
\epsfxsize=6.5cm
\epsffile{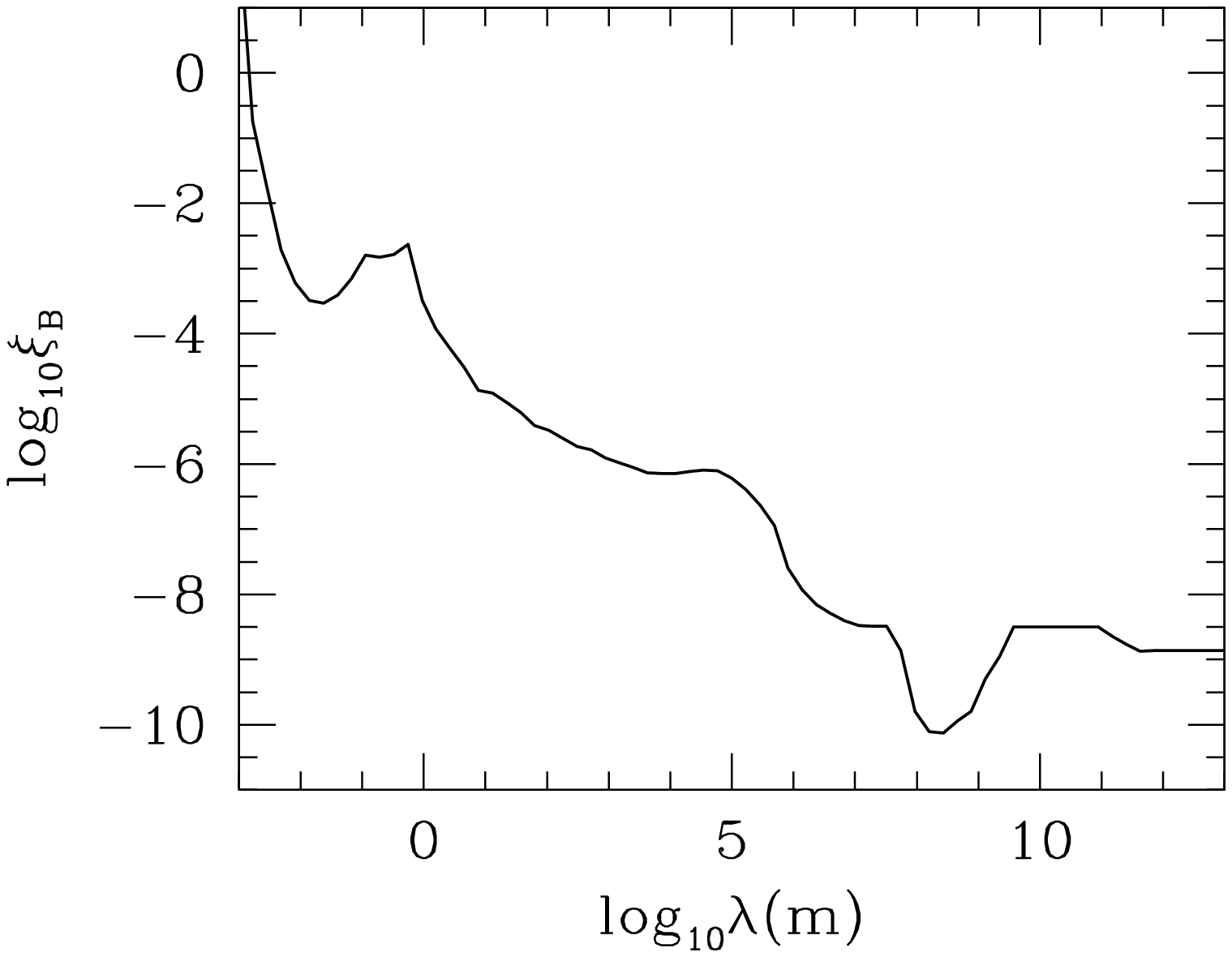}
\mbox{}\\[-.15em]
\hspace*{2em}{\small Composition-dependent experiments\\\hspace{4.4em}
(coupling to baryon number)}
\eminip
\mbox{}\\[1.1em]
\bminip{15cm}
\hspace*{.9em}Figure 4: In each of the diagrams, simplified versions of
Figs. 2.13 and 4.16, for the composition-independent (left) and
-dependent (right) experiments, respectively, of \cite{ft}, for the
non-Newtonian term in the gravitational potential, the curve outlines
the combined boundaries of the disallowed ranges, to the upper side,
plotted against the assumed force-range $\lmd$ in units of meters.

\eminip
\mbox{}\\[1.1em]

The energy eigenvalue $E_i$ will be modified as given by 
\beq
E_{\Phi i}=E_i +<i|\zeta m_{\rm e}\Phi (z)|i>,
\label{3_2}
\eeq
in the first-order perturbation theory with respect to $\Phi$, where the expectation value in the state $i$ is indicated.  By calculating the differences we obtain the level-spacings as given by
\beq
\Delta E_\Phi =\Delta E +\delta \Delta E,\quad\mbox{with}\quad \delta \Delta E=\zeta m_{\rm e} \Phi'(0)a_0 K \sim 10^{-53}\zeta^2, 
\label{3_3}
\eeq
where $a_0=(m_{\rm e}\alpha)^{-1}$ while $K$ is of the order unity, including zero.  As the ratio to the typical level-spacings $\Delta E\sim \alpha^2 m_{\rm e}$, we finally arrive at
\beq
\frac{\delta \Delta E}{\Delta E}\sim 10^{-30},
\label{3_5}
\eeq
for $\zeta^2\sim 10^{-4}$.  In this way we find that the level-spacings of H or  $\bar{\mbox{H}}$ remain virtually unaffected by the NGGF.  This will simplify the analysis in the next subsection.

\subsection{Gravitational redshift from a falling emitter}

A very small number obtained above is  partly due to a quantum mechanical analysis applied to the very small system.  We now discuss the free-fall of $\bar{\mbox{H}}$ or H, basically a classical phenomenon, but combined with the Pound-Rebka type measurement of the gravitational redshift, partly from our own theoretical intrigue.

Let us consider an apparently simplified situation in which an $\bar{\mbox{H}}$ or H falls off freely with the acceleration $\bar{g}=g +\delta g$, where $g$ has an origin in the metric, while $\delta g$ is due to NGGF, so that 
\beq
z_{\mbox{\tiny P}}= -\half \bar{g}t_{\mbox{\tiny P}}^2,
\label{3_6}
\eeq
as shown in Fig. 5, where the subscript P refers to the world point of the falling atom.  Note that the size of $\delta g$ produced by the artificial source mass described before is given by
\beq
\Biggl| \frac{\delta g}{g} \Biggr| \sim 10^{-5}\zeta^2,
\label{3_7}
\eeq
which is $\sim 10^{-9}$ if $\zeta^2\sim 10^{-4}$.  We assume that the acceleration is  constant, just to simplify the analysis.  The falling $\bar{\mbox{H}}$ emits radiation which is received by a receiver R fixed at the height $z=0$ at the world point $\mbox{P}'$.  We are going to apply basically the same analysis used in the Pound-Rebka experiment \cite{mtw}.  

\hspace*{2.6em}
\bminip{13cm}
\mbox{}\\[-7.5em]
\hspace*{-7.5em}
\epsfxsize=9cm
\hspace*{6em}
\epsffile{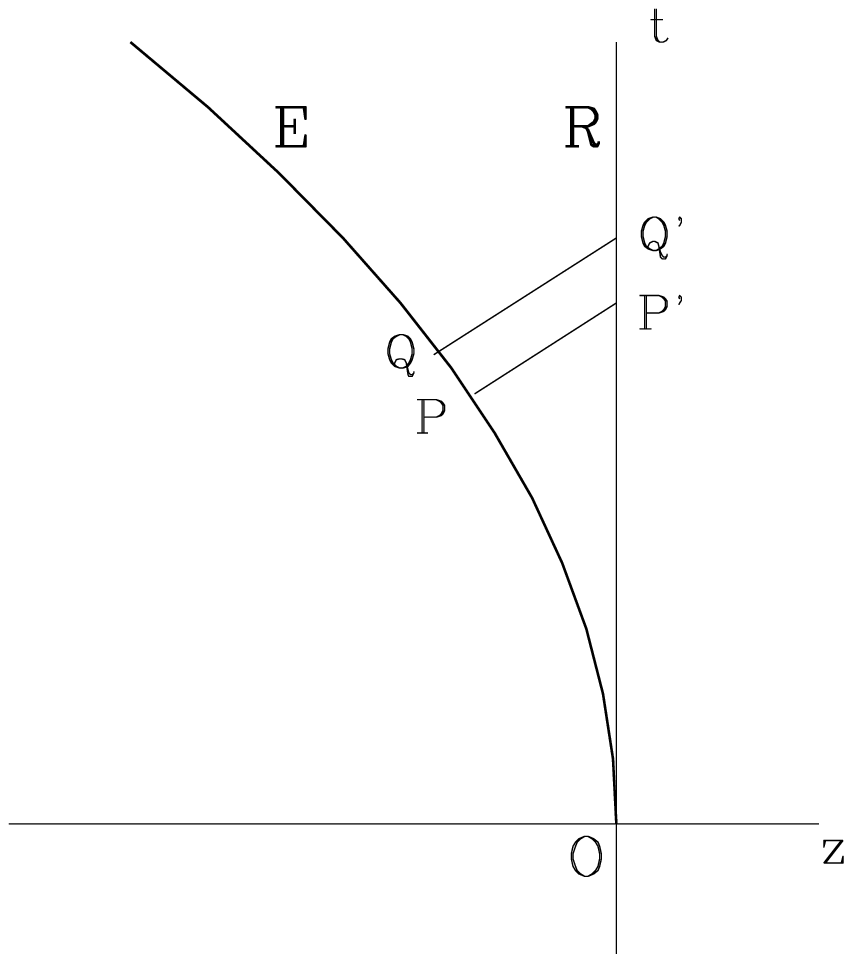}
\mbox{}\\[-2.5em]
Figure 5: E stands for the trajectory of a falling atom emitting radiation at the world point P, received at $\mbox{P}'$ by the receiver R fixed at $z=0$. 
\eminip
\mbox{}\\[1.5em]

We must be careful, however, because this freely-falling emitter with the acceleration $\bar{g}$ does not agree with the local inertial frame in which the frequencies of radiation, emitted and absorbed, are determined on the basis of quantum mechanics in flat spacetime.\footnote{This point was emphasized also in \cite{yfjhf98}.}  In Einstein's sense, the local inertial frame should be defined purely in terms of $g$ without $\delta g$.\footnote{As another detail in this connection, a light-ray propagates through the spacetime endowed with the metric $g$ without affected by NGGF.  The candidate of a vector field does not couple to the  photon due to the charge-conjugation invariance, while the possible coupling to the scalar field has no effect in the geometrical-optics limit \cite{yfms}.}

Taking this dual nature properly into account, we finally obtain
\beq
\frac{\nu_{\mbox{\tiny R}}}{\nu_{\mbox{\tiny E}}}\approx \frac{1+gz_{\mbox{\tiny P}}}{1-\bar{v}}\left( 1-\half \bar{v}^2 +\cdots\right),
\label{3_8}
\eeq
where with $c=1$, $\nu_{\rm R}$ is the frequency measured at the receiver, while $\nu_{\rm E}$ stands for the frequency at the emitter given by one of the level-spacings, for which we exploit the near independence on NGGF, as verified in the preceding subsection.

The term $1+gz_{\mbox{\tiny P}}$ in \reflef{3_8}) is the same as in the ordinary Pound-Rebka result, while the term $(1-\bar{v})^{-1}$ represents the Doppler shift due to the falling of E, where $\bar{v}$ is the downward velocity $\bar{v}=-\bar{g}t_{\mbox{\tiny P}}$ given in terms of $\bar{g}$.  We have another factor which we do not consider any more at this moment.

Roughly speaking, the Doppler shift may determine $\bar{g}$ at relatively earlier times, while the Pound-Rebka type observation determines $g$ at later times.  Suppose we determine or put an upper-bound on $\delta g/g$ to the accuracy of $10^{-9}$.  In view of \reflef{3_7}) we may then have reached nearly the same level of $\zeta^2 \sim 10^{-4}$, as has been derived from other types of estimates in Fig. 4.  But as we mentioned before, those other results might have suffered from considerable uncertainties.  We may also say that the proposed experiment is probably the first to deal with such a simple system like H or $\bar{\mbox {H}}$.  From these points of view, even less accurate results,  say $\delta g/g \sim 10^{-7}$ or $\zeta^2\sim 10^{-2}$, might be still nontrivial at all.  But to go ahead further, more scrutiny is obviously needed both on the theoretical as well as the experimental fronts.


\end{document}